\documentclass[prb,pre,aps,twocolumn,amsmath,amssymb,floatfix,
superscriptaddress]{revtex4-1}
\usepackage[dvips]{graphics}
\usepackage{graphicx}% Include figure files
\usepackage{color}
\usepackage{blindtext}
\definecolor{dred}{rgb}{0.75,0,0}
\usepackage{graphicx}% Include figure files
\usepackage{dcolumn}% Align table columns on decimal point
\usepackage{bm}% bold math
\usepackage{xcolor}
\usepackage{appendix}
\usepackage{hyperref}% add hypertext capabilities
\hypersetup{
 colorlinks=true,
citecolor=red,
 linkcolor=red,
 }
\usepackage{graphicx}% Include figure files
\usepackage{dcolumn}% Align table columns on decimal point
\usepackage{bm}% bold math
\usepackage{xcolor}
\usepackage{hyperref}% add hypertext capabilities
\begin{document}

\preprint{APS/123-QED}

\title{\textcolor{blue}{Ring-localized states, radial aperiodicity and quantum butterflies on a Cayley tree}}

\author{Amrita Mukherjee}
\affiliation{Department of Physics, University of Kalyani, Kalyani,
West Bengal-741 235, India}
\email{amritaphy92@gmail.com}

\author{Atanu Nandy}
\affiliation{Department of Physics, Kulti College, Kulti, Paschim Bardhaman,
West Bengal-713 343, India}
\email{atanunandy1989@gmail.com}

\author{Arunava Chakrabarti}
\affiliation{Department of Physics, Presidency University, 86/1 College Street, Kolkata, West Bengal - 700 073, India}
\email{arunava.physics@presiuniv.ac.in}

\date{\today}% It is always \today, today,

\begin{abstract}
We present an analytical method, based on a real space decimation scheme, to extract the exact eigenvalues of a macroscopically large set of pinned localized excitations in a Cayley tree fractal network. Within a tight binding scheme we exploit the above method to scrutinize the effect of a deterministic deformation of the network, first through a hierarchical distribution in the values of the nearest neighbor hopping integrals, and then through a radial Aubry-Andr\'{e}-Harper quasiperiodic modulation. With increasing generation index, the inflating loop-less tree structure hosts pinned eigenstates on the peripheral sites that spread from the outermost rings into the bulk of the sample, resembling the spread of a forest fire, `lighting up' a predictable set of sites and leaving the rest `un-ignited'. The penetration depth of the envelope of amplitudes can be precisely engineered. The quasiperiodic modulation yields hitherto unreported quantum butterflies, which have further been investigated by calculating the inverse participation ratio for the eigenstates, and a multifractal analysis. The applicability of the scheme to photonic fractal waveguide networks is discussed at the end.

\end{abstract}

\maketitle

%\tableofcontents

\section{Introduction}
\label{Introduction}
The Cayley tree~\cite{cayley} is a paradigmatic example of a tree graph without a closed loop anywhere. The lattice topology permits exact solutions, and has attracted the attention of condensed matter and statistical physics community since long. The canvas of research spans over a wide variety of topics ranging from an exact solution of a localization problem~\cite{abou} and its extensions~\cite{efetov,zirnbauer} to identify the Anderson transition points and the corresponding critical behaviour, modelling dendrimers~\cite{nakano}, studying Heisenberg anti-ferromagnets~\cite{changlani}, or to investigate  the brain-wide neuronal morphology~\cite{lin}, to name a few. 

The literature keeps flourishing, as a hypothesized connection between the Fock space localization  and the occurrence  of single-particle localization on a Bethe lattice ~\cite{altshuler,mirlin,georgeot} rekindle interest in the tree graphs. The idea has subsequently been carried forward to understand the many body localization in a class of spatially extended systems, where localized single particle states exist  and the interaction is short-ranged~\cite{gornyi,basko}.  

The Cayley-tree family offers a unique scope of examining the role of geometry on quantum walks that can, in principle, test the candidature of such branching graphs for a potential application in the transport of quantum information~\cite{li}. A report about the influence of trapped states on a Cayley tree~\cite{mares}, has also been quite revealing in very recent times. The continued interest in unravelling the multifractality of eigenfunction moments~\cite{sonner}, and critical study of the ergodic, non-ergodic extended and localized behaviour of the eigenfunctions~\cite{savitz} on such lattices persist, and with vigor.
%%%%%%%%%%%%%%%%%%%%%%%%%%%%%%%%%%%%%%%%%%%%%%%%%%%%%%%%%
\begin{figure*}[ht]
(a) \includegraphics[width=0.3\columnwidth]{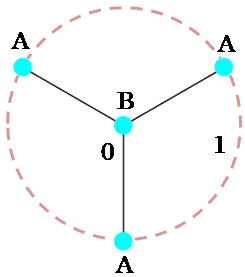}
(b) \includegraphics[width=0.45\columnwidth]{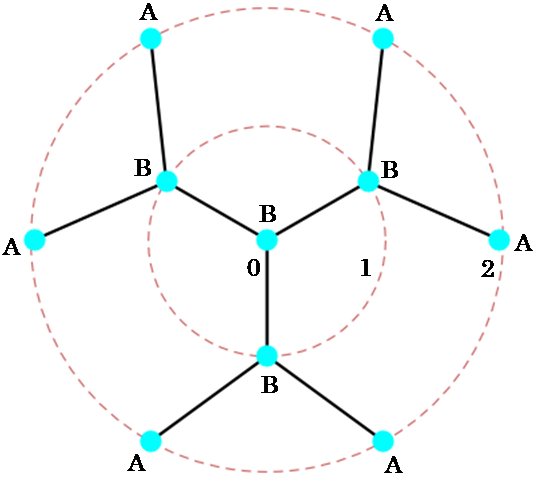}
(c) \includegraphics[width=0.55\columnwidth]{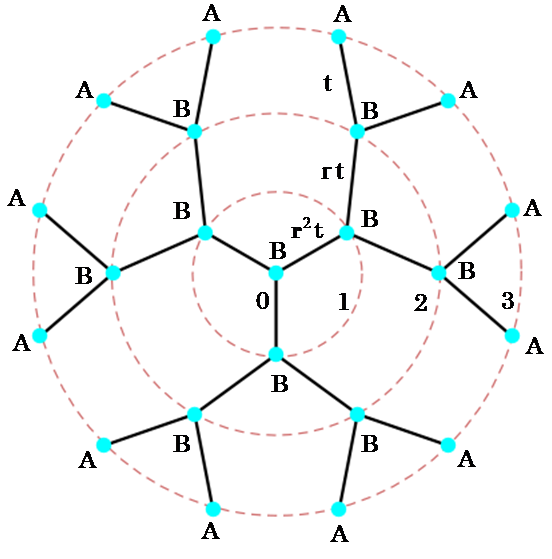}
	\caption{(Color online.) The schematic diagram of the first three generations of a $b=2$ Cayley tree network. Shells (dashed red circles) are numbered. The outermost sites having coordination number two are marked as $A$. All other sites, including the seed at the centre, have three nearest neighbors, and are named $B$.}
	\label{tree1}
\end{figure*}
%%%%%%%%%%%%%%%%%%%%%%%%%%%%%%%%%%%%%%%%%%%%%%%%%%%%%%%%%

However, a subtle issue apparently gets overlooked. For fractals, tree-like or of any other topology, the single particle states are, in general, localized. Is it possible to know the {\it precise} energy eigenvalues for states localized in a fractal, branching or non-branching, even when the fractal grows indefinitely large? A direct diagonalization of a finite system of course yields a set of such localized state-eigenvalues, but with the growth of the system, the values obtained with a smaller generation, in general, slip away, and a new larger set is obtained. Can we have a prescription that enables one to extract at least a subset of eigenvalues for a deterministic, but non-translationally invariant self-similar fractalline  structure, however large it may be? This is the question that motivates us to undertake a deeper study of the Cayley tree network.

We use a tight binding Hamiltonian for studying a $b=2$ (the branching number) Cayley tree network (Fig.~\ref{tree1}) and prescribe a scheme that works out the {\it exact}, tailor-made  eigenvalues corresponding to a class of pinned localized single particle states on such a tree graph at any generation. The method outlined here works for a wide variety of disorder introduced in the structure. In this particular communication we address the case of spinless fermions on a Cayley tree graph. Two separate cases are examined. First, we investigate the case where the inter-shell (explained in the text later) overlap (hopping) integrals follow a {\it hierarchical distribution}. The distribution is explained in Fig.~\ref{tree1}(c) (the $G_2$ generation). Second, we introduce, for the first time, a {\it radial aperiodicity} in the distribution of the inter-shell hopping. Naively speaking, such variations in the hopping term imply deterministic deformation of the graph, as the amplitude of hopping can be related to the topological distance between the sites~\cite{eckstein}. The Cayley tree that we study here, can thus be treated as a tree-graph with a sequentially {\it squeezed} or {\it expanding} branch lengths, depending on whether the hierarchy parameter $R$ is greater than, or less than unity. To the best of our knowledge, such an issue has remained unaddressed so far.

It turns out that, just by simple observation and using a recursive decimation scheme, it is possible to extract, in a completely analytical way, a whole subset of eigenvalues corresponding to localized amplitudes {\it pinned} on predictable subsets of the vertices. As we extract the eigenvalues from deeper and deeper recursion levels, the amplitudes of the eigenstates start penetrating the tree from the outermost branch-ends (forming ring-like structure) into the bulk, exactly resembling the spread of a forest-fire. The distribution can be cut off at any desired level (depth) at will, leaving the rest of the graph including the seed vertex in {\it dark}, that is, with zero amplitudes.

The results presented here are interesting in their own merits. With a hierarchically designed hopping amplitude profile, the pinned localized states, residing on the rims of the shells exhibit an interesting distribution in the energy eigenvalues, as the hierarchy parameter is varied. The pinned localized states, residing on the rings are found to span an energy interval in the neighborhood of the spectral centre (here $E=0$) only for a selected window of the hierarchy parameter, indicating a hitherto unemphasized spectral behaviour of such an open graph. 

The quasiperiodic Aubry-Andr\'{e}-Harper modulation~\cite{aubry,harper}, introduced in the distribution of the inter-shell hopping integrals, spreading radially {\it inward} from `leaf to root' on the other hand, implies a different kind of deterministic deformation  in the branching pattern of the tree. The spectrum is seen to give rise to what we name here as a {\it Cayley-Hofstadter butterfly} (CHB) family that exhibits a multifractal character and a seemingly possible localization-delocalization transition of single particle states as the strength of the aperiodic modulation is altered. The observations are supported by a multifractal analysis and a detailed calculation of the inverse participation ratio (IPR).

Finally, we suggest an experiment with single mode photonic waveguides crafted in a Cayley tree pattern. The partially lit up tree, and the forest fire-like propagation of photonic modes can, in principle, be `engineered' at appropriate frequencies. Such frequencies are derived exactly following our prescription and using the classical wave equation, as discussed.

In the subsequent sections II, III and IV, we discuss the basic methodology formulated in terms of non-interacting spinless fermions, introduce and analyse the case of a radial aperiodicity in the distribution of the hopping integrals, and provide a multifractal analysis respectively. In section V we propose a photonic equivalent of the present study that can inspire experiments, and in section VI we draw our conclusion.
%%%%%%%%%%%%%%%%%%%%%%%%%%%%%%%%%%%%%%%%%%%%%%%%%%%%%%%%%
\begin{figure*}[ht]
(a) \includegraphics[width=0.6\columnwidth]{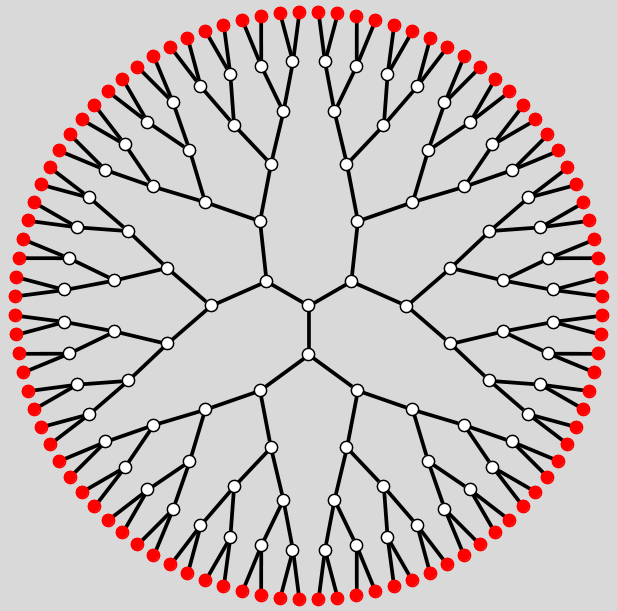}
(b) \includegraphics[width=0.6\columnwidth]{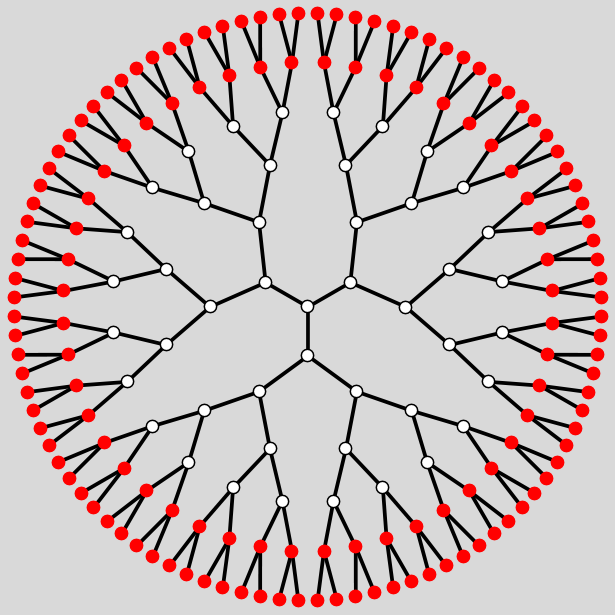}
(c) \includegraphics[width=0.6\columnwidth]{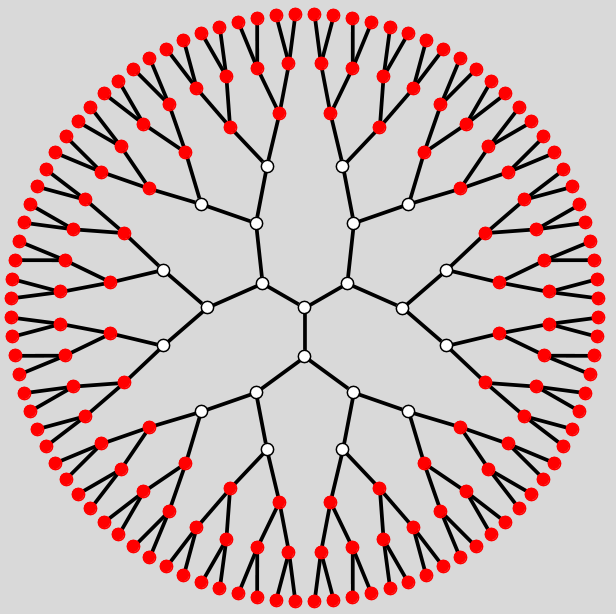}
	\caption{(Color online.) Forest fire propagation of non-zero amplitudes on a $3^{rd}$ generation $b=2$ Cayley tree. The appearance of the amplitudes (red dots) can be engineered, as explained in the text.}
	\label{forestfire}
\end{figure*}
%%%%%%%%%%%%%%%%%%%%%%%%%%%%%%%%%%%%%%%%%%%%%%%%%%%%%%%%%
\section{Extracting the ring localized states}
\subsection{The Hamiltonian}

A Cayley tree with a coordination number $z$ and having $S$ number of `shells' (shown as dashed rings in Fig.~\ref{forestfire}) allows an exact evaluation of the number of atomic sites in any finite generation~\cite{yorikawa}. In the $s$-th shell of the tree one finds $N_s = z (z-1)^{s-1}$ sites. $s$ runs from $1$ to $S$. The number of bulk or `interior' sites of an $S$-shell tree is $N_{bulk} = [N_{S}-2]/(z-2)$, and the total number of sites for an $S$-shell tree is, $N=[(z-1) N_S -2]/(z-2)$.
We shall use the usual form of the tight binding Hamiltonian, describing nearest neighbor hopping on a Cayley tree depicted in Fig.~\ref{tree1}. The Hamiltonian is,  
\begin{equation}
    H_S = \sum_{s,i \in s} \epsilon_{i,s} c_{i,s}^\dag c_{i,s} +  
   \sum_{<s,s'>} \sum_{i\in s,j\in s' } t_{i, s}^{j,s'} c_{j,s'}^\dag c_{i,s}  + h.c
    \label{ham}
\end{equation}
$\epsilon_{i,s}$ is the on-site potential at the $i$-th site on the $s$-th shell.  
The hopping integral $t_{i,s}^{j,s'}$ is non-zero only between the connected nodes on the nearest neighboring shells $s$ and $s'$. $i$ and $j$ refer to the nearest neighbors residing on two nearest neighboring shells only, shown connected by the bonds (branches), drawn by black lines in Fig.~\ref{tree1}. 

We shall use the difference equation satisfied by the amplitudes of the wave function at every node $(i,s)$, written as, 
\begin{equation}
    (E-\epsilon_{i,s}) \psi_{i,s} = \sum_{j} \left [ t_{j,s}^{j,s+1} \psi_{j,s+1} + t_{j,s}^{j,s-1} \psi_{j,s-1} \right ]
    \label{diff}
\end{equation}
where, $j$ stands for the indices of sites on the $(s+1)$-th or the $(s-1)$-th shell that are nearest neighbors to the $i$-th site on shell number $s$.
\subsection{A hierarchically deformed tree}
%%%%%%%%%%%%%%%%%%%%%%%%%%%%%%%%%%%%%%%%%%%%%%%%%%%%%%%%%
\begin{figure}[ht]
 \includegraphics[width=0.6\columnwidth]{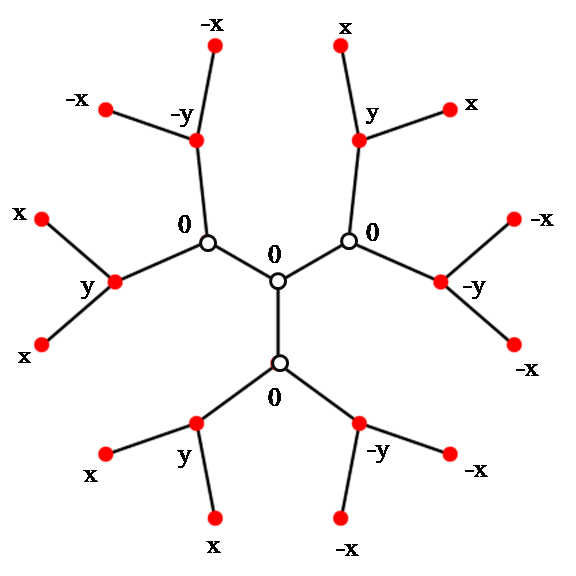}
	\caption{(Color online.) Pictorial presentation of amplitude distribution for a $3^{rd}$ generation cayley tree network at energy eigenstate $E=\sqrt{2}$. Here, $x=1$, and $y=\sqrt{2}$. }
	\label{amplitree}
\end{figure}
%%%%%%%%%%%%%%%%%%%%%%%%%%%%%%%%%%%%%%%%%%%%%%%%%%%%%%%%%
We explain the scheme here. The hopping integrals on an $S$-shell Cayley tree have been chosen to follow a hierarchical pattern, starting at a value `$t$' connecting the sites residing on the $S$-th (the outermost) and the $(S-1)$-th shells, and subsequently having a shell-to-shell distribution of values (propagating inward) following the sequence $rt$, $r^2t$, $....r^{S-1}t$ down to the seed (red site in Fig.~\ref{tree1}). The values of the on-site potentials are set as $\epsilon_A$ for the outermost $A$ sites having a coordination number one and residing  on the rim of the terminal shell $S$. The bulk sites sitting on the rims of all other remaining shells, including the seed, have a coordination number equal to three each, and are assigned a potential $\epsilon_B$. Appreciating that the value of the inter-shell hopping integrals can be related to the topological (radial) distance between the branches, one can take $r > 1$ or $ r < 1$ to mimic a deterministic, radially squeezed or stretched Cayley tree graph.

Let us fix $\epsilon_A=\epsilon_B=0$ to begin with. It is easy to see that, if we choose $E=0$, then by setting the wavefunction amplitudes $\psi_{j,S}$ at the $A$ sites on the rim of the outermost $S$-th shell in a periodic sequence of $\pm 1$ along the perimeter,  Eq.~\eqref{diff} is trivially satisfied with the remaining amplitudes $\psi_{j,s}$ ($s \ne S$) at {\it all} the remaining shells, including the seed, equal to zero. This is illustrated by the red dots on the outermost periphery in Fig.~\ref{forestfire}(a) representing the non-zero amplitudes, pinned on a ring, and the white dots representing the sites at which the amplitude of the wavefunction is zero. The amplitude profile at $E=0$, shown by the series of red dots in Fig.~\ref{forestfire}(a) resembles the outermost line of a simulated {\it forest fire}, as we may call it.
%%%%%%%%%%%%%%%%%%%%%%%%%%%%%%%%%%%%%%%%%%%%%%%%%%%%%%%%%
\begin{figure*}[ht]
(a)	\includegraphics[width=0.5\columnwidth]{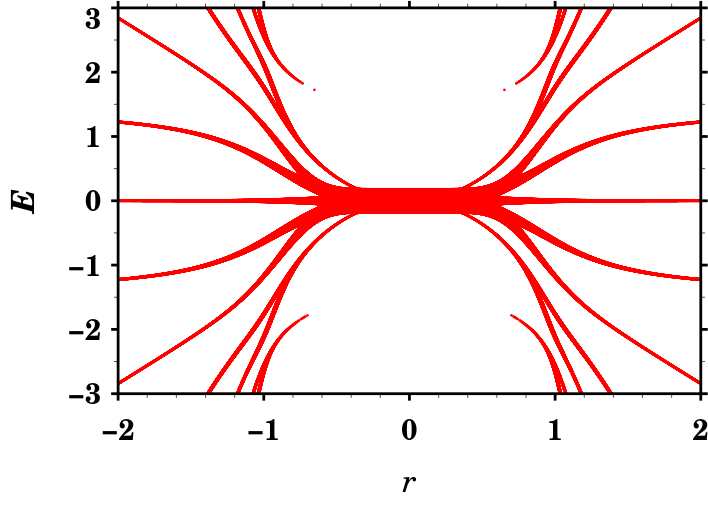}
(b)	\includegraphics[width=0.5\columnwidth]{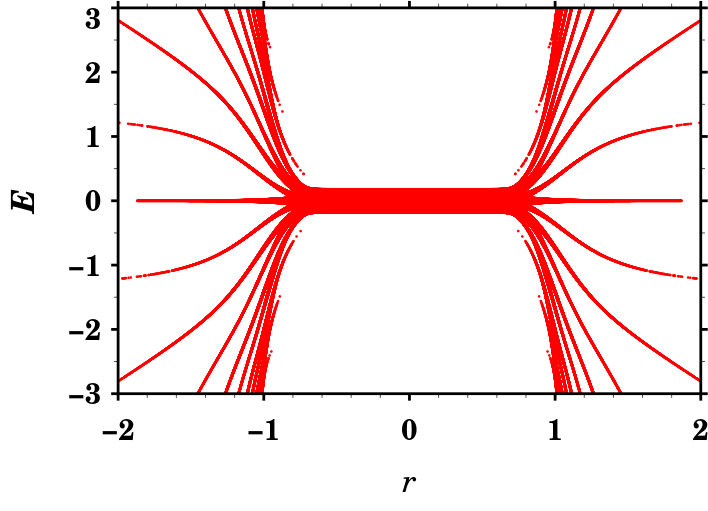}\\
(c)	\includegraphics[width=0.5\columnwidth]{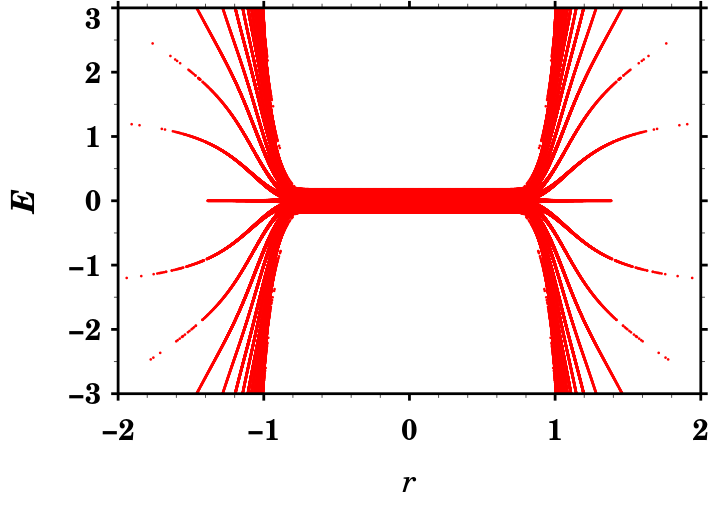}
(d)	\includegraphics[width=0.5\columnwidth]{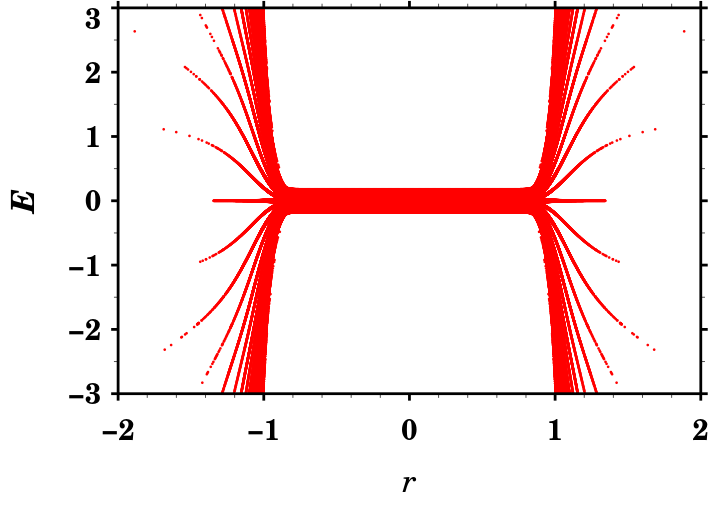}
	\caption{(Color online.) Distribution of energy for a hierarchical leaf-to-root distribution of hoppings on  (a) $5^{th}$ generation, (b) $10^{th}$ generation, (c) $15^{th}$ generation, and (d) $20^{th}$ generation $b=2$ Cayley tree.}
	\label{hier}
\end{figure*}
%%%%%%%%%%%%%%%%%%%%%%%%%%%%%%%%%%%%%%%%%%%%%%%%%%%%%%%%%
Eq.~\eqref{diff} can now easily be exploited to decimate the amplitudes $\psi_{j,S}, j\in A$ on the outermost shell $S$. This scales the original tree to a `renormalized' version where the $(S-1)$-th shell in the original structure now takes up the role of the `new outermost' shell. The last hopping integral now `connects' the shells $(S-2)$ and $(S-1)$, and is equal to $rt$ in this case. The new outermost sites are now the `converted' $B$-sites on the $(S-1)$-th shell at the bare length scale. The potential on this new outermost shell and the hopping integral connecting the first pair of outermost shells are given by, 
\begin{equation}
\epsilon_{A,1} = \epsilon_{B,0} + \frac{2 t^2}{E - \epsilon_{A,0}} 
\label{decimation-1}
\end{equation}
where the subscripts $1$ and $0$ refer to the number of decimation steps. Obviously, $\epsilon_{A(B),0} = \epsilon_{A(B)}$.
%%%%%%%%%%%%%%%%%%%%%%%%%%%%%%%%%%%%%%%%%%%%%%%%%%%%%%%%%
\begin{figure*}[ht]
%(a)\includegraphics[width=0.5\columnwidth]{cayley-aubry.png}\\
(a)\includegraphics[width=0.5\columnwidth]{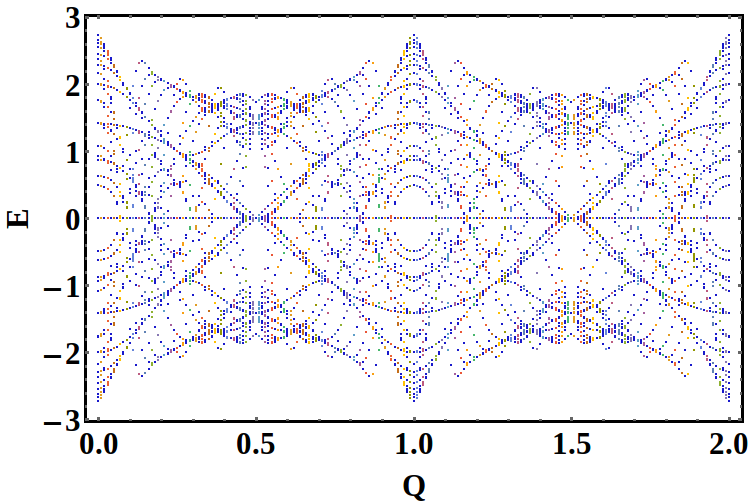}
(b)\includegraphics[width=0.5\columnwidth]{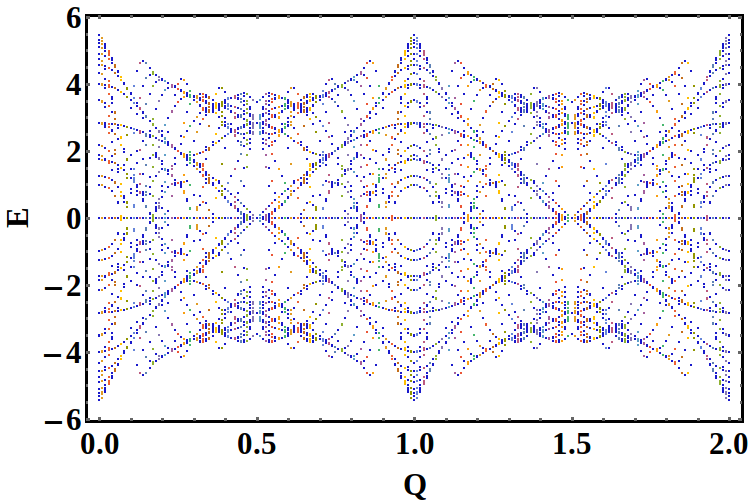}\\
(c)\includegraphics[width=0.5\columnwidth]{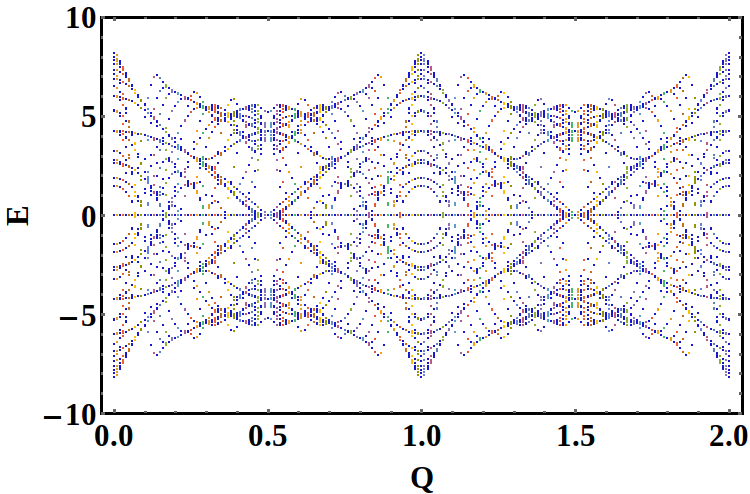}
(d)\includegraphics[width=0.5\columnwidth]{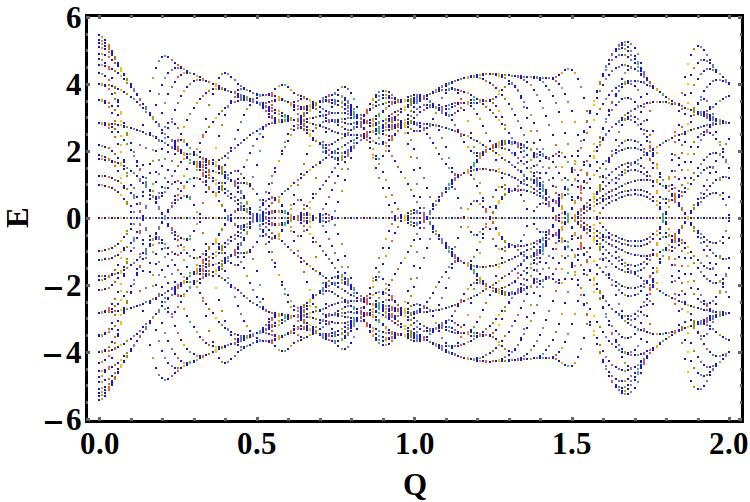}
	\caption{(Color online.) The distribution of allowed eigenstates against $Q$ for a Cayley tree with off-diagonal AAH model of $8$-th generation for three different values, (a) $\lambda=1$, (b) $\lambda=2$, (c) $\lambda=3$. (d) shows the distortion of the Cayley-Hofstadter butterfly at $\nu=0.8$ for $\lambda=2$.}
	\label{butterfly}
\end{figure*}
%%%%%%%%%%%%%%%%%%%%%%%%%%%%%%%%%%%%%%%%%%%%%%%%%%%%%%%%%
One can now extract eigenvalues for the second set of pinned states by solving the equation $E=\epsilon_{A,1}$ which yields 
\begin{equation}
    E = \frac{\epsilon_A + \epsilon_B \pm \sqrt{(\epsilon_A - \epsilon_B)^2 + 8t^2}}{2}
\end{equation}
The corresponding {\it non-zero} amplitudes now span the two consecutive outermost rings, decorated with red dots in Fig.~\ref{forestfire}(b). With $\epsilon_A=\epsilon_B=0$, and $t=r=1$, the energy eigenvalues corresponding to the ring-localized states are $E = \pm \sqrt{2}$. The amplitudes are $\pm 1$ and $\pm \sqrt{2}$, and the distribution is depicted in Fig.~\ref{amplitree}. Beyond the $(x,x,y)$ or the $(-x,-x,-y)$ clusters the remainder of the graph hosts only {\it zero} amplitudes, as explained above. A further pattern of distribution of amplitudes, and the penetration of the forest fire on a two times decimated tree is illustrated in Fig.~\ref{forestfire}(c). The states are pinned on the first three outermost rings now, at $E=0$, $\pm 2$. To avoid a cumbersome representation, we omit the precise numerical values in the figure, and just paint the sites in red.

This decimation can be continued sequentially, beginning at the last, $S$-th shell of any arbitrarily large tree, and `folding' the shells (by decimating the vertices on the periphery) one by one. A general expression for the renormalized on-site potential $A$ at an $(n+1)$-th level, after a decimation is done $n$ times, is written as, 
\begin{equation}
    \epsilon_{A,n+1} = \epsilon_{B,0} + \frac{2 r^{2n} t^2}{E-\epsilon_{A,n}}
    \label{recursion}
\end{equation}

In Fig.~\ref{hier} we plot the distribution of the pinned state-eigenvalues as the hierarchy
parameter $r$ is varied. A dense cluster of pinned state-eigenvalues are always present around the centre of the spectrum $E=0$, but with the hierarchy parameter lying only within a special window. With increasing generation this central `band' seems to be clamped between $r=\pm 1$. Beyond $r=\pm1$ the pinned localized states exhibit an interesting fanning out effect, with the fanning branches converge towards $r=\pm 1$ boundaries and climbing along the energy axis. However, the spectrum always remain bounded.

\section{Radial aperiodicity and the trapped eigenstates}
We introduce a radial aperiodicity in the distribution of the hopping amplitudes to see the effect of a `deterministic disorder' on the energy spectrum offered by a Cayley tree. The hopping modulation follows a modification~\cite{sankar1,sankar2} over the celebrated Aubry-Andr\'{e}-Harper (AAH) model~\cite{aubry,harper},  and is defined as, 
\begin{equation}
    t_n = \lambda \cos (\pi Q n^\nu)
\label{aubry}    
\end{equation}
Here, $\lambda$ is the strength of the modulation, and a choice of $Q$ as an irrational number, say the golden ratio $(\sqrt{5}+1)/2$ (as we have taken here later) brings in the true incommensurate limit. The index $n$ runs from $S$ to $1$ over the shells, starting from the outermost shell $S$ to the first shell $1$, thereby implying a `leaf to root' variation in the hopping strength. As is well known, the above form of the modulation shows mobility edges even in one dimension~\cite{sankar1,sankar2}. The key factor is the `slowness parameter' $\nu$, that creates non-trivial changes over the canonical results of the celebrated AAH model~\cite{aubry,harper}, where no mobility edge exists, and a metal-insulator transition is seen only in the parameter space.
%%%%%%%%%%%%%%%%%%%%%%%%%%%%%%%%%%%%%%%%%%%%%%%%%%%%%%%%
\begin{figure}[ht]
(a)\includegraphics[width=.5\columnwidth]{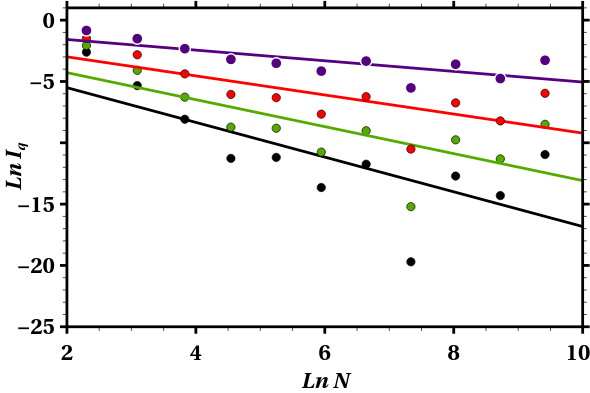}\\
(b)\includegraphics[width=.5\columnwidth]{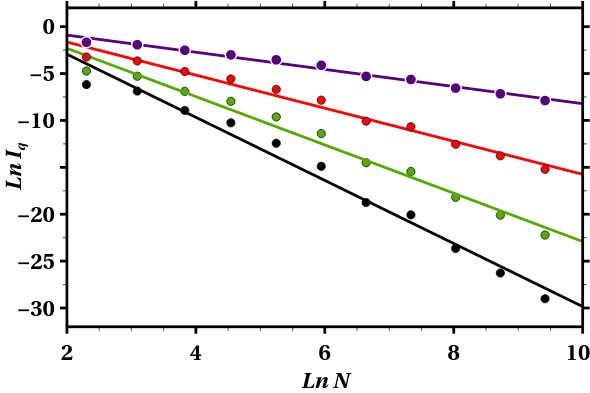}\\
(c) \includegraphics[width=.5\columnwidth]{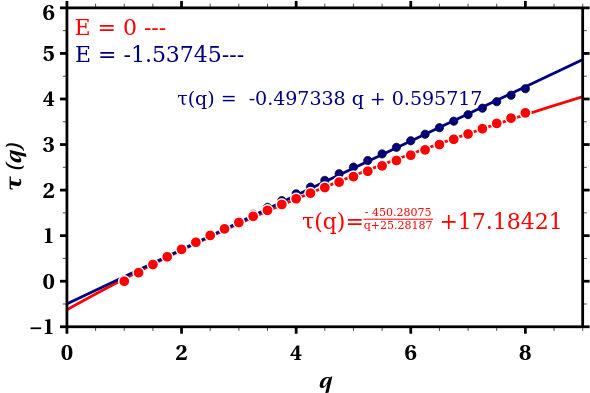}
	\caption{(Color online.) (a) and (b) show about the distribution of $\ln{I_q}$ vs. $\ln{N}$ for two different energies (a) $E =
0$, (b) $E = -1.53745$ with the number of sites $N$ in the system varying from $10$ to
$12286$. Here the parameter $q$ takes four different values 2 (violet), 3 (red), 4
(green), and 5 (black). 
(c) Plot of multifractal scaling exponent $\tau(q) \equiv -\frac{\ln{I_q}}{\ln{N}}$ with q for a system size with 382
atoms. Here, we fix $\lambda=3$, $\nu=1$, and $Q=\frac{\sqrt{5}+1}{2}$.}
\label{multifractal}
\end{figure}
%%%%%%%%%%%%%%%%%%%%%%%%%%%%%%%%%%%%%%%%%%%%%%%%%%%%%%%%%
The eigenvalue spectra of a Cayley tree network with a radial aperiodicity, are plotted in Fig.~\ref{butterfly} as the parameter $Q$ is varied. As the slowness parameter $\nu$ is set equal to unity, reflecting a clean AAH variation, the CHB structures are apparent with the fractal nature revealed in a conspicuous manner. With $\nu$ deviation from unity, the slow modulation in the hopping distribution sets in and a {\it death} of the quantum butterflies is signalled. 

\section{Multifractal analysis}
The multifractality of the eigenspectrum of the Cayley tree network with radial aperiodicity in the off-diagonal term of the Hamiltonian can be characterized following a standard prescription~\cite{evers2008}. Multifractality of the wave functions demands that the eigenstates exhibit strong fluctuations at the critical point~\cite{janssen,aoki} and this is found to be the general feature of eigenstates at the mobility edges for Anderson transitions~\cite{evers2008}.
%%%%%%%%%%%%%%%%%%%%%%%%%%%%%%%%%%%%%%%%%%%%%%%%%%%%%%%%%
\begin{figure}[ht]
(a) \includegraphics[width=0.5\columnwidth]{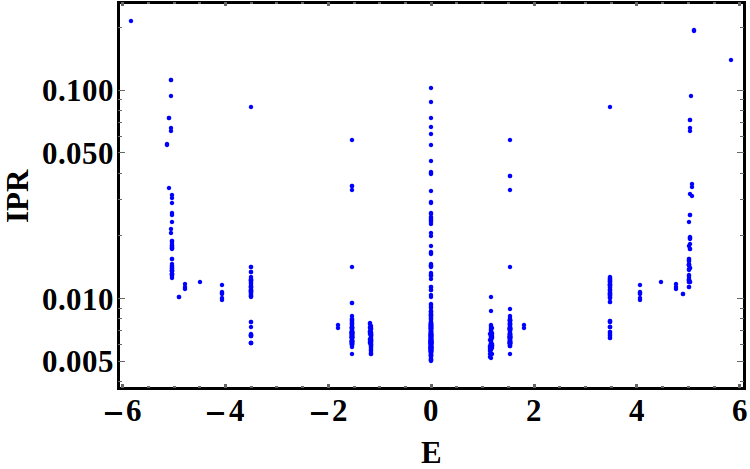} \\
(b) \includegraphics[width=0.5\columnwidth]{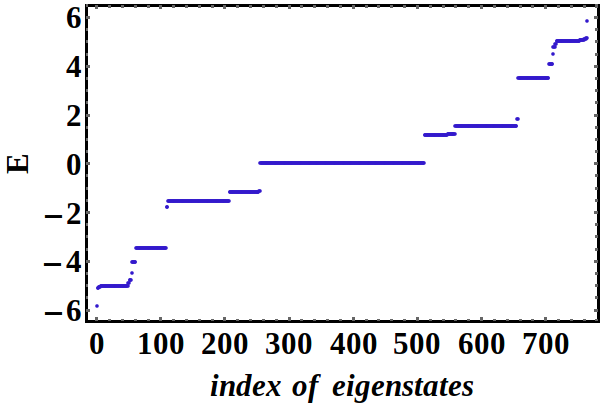}
%(b) \includegraphics[width=0.5\columnwidth]{trans-aubry-lambda-3-gen-5.png}
\caption{(Color online.) (a) Plot of Inverse participation ratio for an $8$th generation Cayley tree as a function of energy. A radial aperiodicity is induced through the AAH model which is incorporated in the off-diagonal hopping integrals.(b) The distribution of degeneracy of the eigenstates. In both (a) and (b) we have set $Q=(\sqrt{5}+1)/2$,  $\lambda=3$, and $\nu=1$. $\epsilon=0$ everywhere, as before.}
	\label{ipr-lambda-3}
\end{figure}
%%%%%%%%%%%%%%%%%%%%%%%%%%%%%%%%%%%%%%%%%%%%%%%%%%%%%%%%%

The classic signature of the multifractal systems is captured in a continuous set of exponents describing the scaling of the moments of a probability distribution.
The multifractal behavior can be symbolized using the generalized inverse participation ratio, defined as
\begin{equation}
\mathcal{I}_{q} = \int d \mathbf{r} | \psi (\mathbf{r})|^{2q}
\label{mfaeq1}
\end{equation}
Here the integration is carried over the real lattice space. At the critical point, $\mathcal{I}_q$ reveals an anomalous scaling behavior with the system dimension $N$, viz,  $\mathcal{I}_q \sim N^{-\tau_{q}}$. The factor $\tau_{q}$ describes the set of exponents that basically contains substantial information regarding the critical nature of the states. 

It is customary to introduce a fractal dimension $D_q$ via $\tau_q = D_q (q-1)$. Here, the necessary constraint is the normalization of wave function, which requires $\mathcal{I}_1 = 1$. For structures in $d$ dimensions, $D_q = d$ signifies a metallic state, \textit{i.e.}, the associated spectrum turns out to be continuous leading to the existence of extended eigenstates.
On the other hand, $D_q = 0
$ is a typical signature of insulating or non-transmitting nature of the state (point spectrum for localized states).
The intermediate phase is characterized via a nonlinear functional dependence of $D_q$ on $q$. This is a clear indication of the multifractal nature of the wave function at the critical point~\cite{evers2008}.
Fig.~\ref{multifractal} demonstrates the multifractality observed in the energy spectrum of a Cayley tree with radial aperiodic distribution of the hopping integral. We have examined the multifractal behaviour for $\nu=1$, as well as for its other fractional values. The global character of multifractality is similar for different values of $\nu$, and hence we present the results only fo $\nu=1$ to save space. The  multifractal analysis adopted here unravels the non-trivial nature of underlying quantum phases associated with such eigenstates. It is needless to say that this well-known and standard method has already been used in studying several other disordered networks~\cite{rudo,luck,hiramoto,liu}.

We simultaneously examine two characteristically different eigenstates that always belong to the spectrum as the system grows larger. The two distinct kinds of states considered here correspond to $E=0$ and $E=-1.53745$ and belong to the high and the low IPR regimes respectively. As expected, these states have a considerable difference in the localization area as it is evident from the IPR plot shown in Fig.~\ref{ipr-lambda-3}(a). We set the value of $Q$ at its  aperiodic limit \textit{i.e.}, we set $Q=(\sqrt{5}+1)/2$ and the slowness index is set at $\nu =1$. 
%%%%%%%%%%%%%%%%%%%%%%%%%%%%%%%%%%%%%%%%%%%%%%%%%%%%%%%%%%%%%

Speaking in terms of the characteristic difference between the two modes, the plot of generalized IPR also highlights the distinction between the two quantum phases. Fig.~\ref{multifractal}(c) narrates the dependence of generalized IPR on $q$. It is clear from the profile that for the $E=0$ state the functional dependence is non-linear in the way, 
\begin{equation}
\tau_q = \frac{a}{q+b}+c
\end{equation}
A best fit yields $a=-450.28075$, $b=25.28187$, and $c=17.18421$.
For $E=-1.53745$, the dependence follows a straightforward linear relationship $\tau_q = a q + b$ with $a=-0.497338$, and $b=0.595717$.  Hence, the metallic signature is apparent for the mode $E=-1.53745$ and on the contrary the prominent non-linear change of $\tau_q$ indicates a  {\it critical} behavior of the eigenstate at $E=0$.

We conclude this subsection with a discussion on the degeneracy of the eigenstates. Almost all the eigenstates exhibit degeneracy at a macroscopic level, and display a multitude of localization areas. That is, for each such degenerate modes the envelope of the eigenfunction encompasses a varied number of lattice points,  depending on the degree of degeneracy. The degeneracy is explicitly demonstrated in in Fig.~\ref{ipr-lambda-3}(b), where the distribution of energy eigenvalues is plotted against the site index. The figure exhibits a staircase pattern, showing the largest degeneracy at $E=0$, while the degeneracy gets reduced towards the ends of the spectrum, and this feature is universal, that is, holds for any value of the modulation strength $\lambda$. Of course, one can locate the eigenvalues in the spectrum which turn out to be non-degenerate, and are represented by isolated dots.

%\begin{figure}[ht]
%\includegraphics[width=0.8\columnwidth]{cayley-aubry.png}
%(a) \includegraphics[width=0.5\columnwidth]{cayley-aubry-energy-lambda-1.png}
%(b) \includegraphics[width=0.5\columnwidth]{cayley-aubry-energy-lambda-2.png}
%(c) \includegraphics[width=0.5\columnwidth]{cayley-aubry-energy-lambda-3.png}\\
%(a) \includegraphics[width=0.5\columnwidth]{cayley-aubryhop-lambda-1.png} \\
%(b) \includegraphics[width=0.5\columnwidth]{cayley-aubryhop-lambda-2.png} \\
%(c) \includegraphics[width=0.5\columnwidth]{cayley-aubryhop-lambda-3.png} 
%	\caption{(Color online.) Distribution of energy eigenvalues against the index of eigenstates for an 8th generation Cayley tree network, bringing out the degeneracy staircase when, (a) $\lambda=1$, (b) $\lambda=2$, and (c) $\lambda=3$. 
%	 We have taken $Q=(\sqrt{5}+1)/2$, and $\nu=1$.}
%	\label{eng}
%\end{figure}
%%%%%%%%%%%%%%%%%%%%%%%%%%%%%%%%%%%%%%%%%%%%%%%%%%%%%%%%%
\section{Extension to photonics}
The methodology proposed here to extract the ring states localized around the peripheral sites, penetrating gradually into the bulk, but excluding a predictably selected set of site (thereby mimicking a forest fire distribution), can be easily extended to a Cayley-tree photonic waveguide network, where each branch is a single mode photonic waveguide. The grafted photonic lattice can serve as the testing ground for such a theory, as presented so far. 

The wave propagation can be described using the variation of the amplitude of the wave excitation between the pair of nodes $(i,j)$ in the waveguide, which is given by~\cite{pingsheng},
\begin{equation}
 \psi_{ij}(x) = \psi_i \frac{\sin[(l_{ij}-x)k]}{\sin(l_{ij} k)} +
\psi_j \frac{\sin(x k)}{\sin(l_{ij} k)}
\label{ampli}
   \end{equation}
Consideration of the flux conservation at a node~\cite{alexander} maps the wave equation into a `tight binding', `difference equation analogue', viz, 
\begin{equation}
-\psi_i \sum_j \cot\theta_{ij} + \sum_j \csc \theta_{ij} \psi_j =0
\label{light}
\end{equation}
where, $\theta_{ij}=k l_{ij}$. $l_{ij}$ is the length of a waveguide, which we may select as constant, to implement in the easiest of experiments. This choice makes the phase acquired by the wave travelling in a segment uniform, that is, $\theta_{ij}=\theta$ for all. The summation over $j$ counts the number of vertices connected to the $i$th node. The identification of $\epsilon_A \equiv \cot \theta$, and $\epsilon_B \equiv 3 \cot \theta$ is obvious, when we form a Cayley tree waveguide network with the refractive index and the length $l_{ij}$ same for every waveguide segment. 

Once this is accomplished, the exact solution of the wave equations, using the same scheme designed to address the spinless fermions on such a network, reveals a forest-fire like spread of {\it illumination}, burning bright at the outermost vertices of an arbitrarily large tree, and gradually penetrating the bulk. The degree of spread and the hierarchy of wave-frequencies can be exactly evaluated. for example, to glow only the $A$ vertices in the  outermost ring, one needs to `inject' light of wave vector $k=\pi/2l$. A one step decimation yields the result that, to ignite the two consecutive rings starting at the outermost ring,  $k=(1/l) \cos^{-1}(\sqrt{2/3})$. One can easily extend the scheme now to extract the precise frequencies needed to achieve the selected glow-pattern on such a waveguide network. The frequencies turn out to be densely packed as the tree grows larger and larger in size. The detailed analysis will be published elsewhere. The idea and the results can, to our understanding, add to the list of curiosities of the photonic waveguide experimentalists. 

\section{Conclusion}
\label{summary}
We have revisited the problem of localization on a Cayley tree network from a different perspective. We begin with a simple geometrical construction of amplitudes of the wavefunction pinned on the peripheral sites, and then used a real space decimation idea to extract a whole family of such {\it pinned states} which flow from the outer leaves to the root, but can never be non-zero beyond a certain scale. This pattern resemble a forest fire. We then introduce, for the first time to the best of our knowledge, a radial aperiodicity in the distribution of the hopping amplitudes - a choice that essentially mimics a tree where the branches are {\it deterministically distorted}. The Hofstadter butterflies show up when we introduce the aperiodicity following an Aubry-Andr\'{e}-Harper pattern. A thorough multifractal analysis and inverse participation ratio are provided to explore the fractality of the spectrum, and to study any possible crossover from low to high localization areas. Finally, the possibility of designing single mode photonic waveguides is pointed out. If possible, such a waveguide network should be able to prove with a unique, engineered `selectively ignited' Cayley tree structure. A detailed analysis of this issue will be presented elsewhere.
%\begin{figure*}[ht]
%\includegraphics[width=0.8\columnwidth]{cayley-aubry.png}
%(a) \includegraphics[width=0.5\columnwidth]{cayley-hopaubry-ipr.png}
%	\caption{(Color online.) Schematic distribution of inverse participation ratio against disorder strength at $E=0$. We have taken $Q=(\sqrt{5}+1)/2$.}

%	\label{eng}
%\end{figure*}
%%%%%%%%%%%%%%%%%%%%%%%%%%%%%%%%%%%%%%%%%%%%%%%%%%%%%%%%%

%\begin{figure*}[ht]
%\includegraphics[width=0.8\columnwidth]{cayley-aubry.png}
 %\includegraphics[width=0.8\columnwidth]{aubry-trans-cayley.png}
%(b) \includegraphics[width=0.5\columnwidth]{trans-aubry-lambda-3-gen-5.png}
	%\caption{(Color online.) Plot of transmission profile as a function of energy for various generation($N$) of Cayley tree network. $N=1$, $N=2$, $N=3$ are defined by red color, blue color and green color respectively. A radial aperiodicity is induced through the AAH model which is incorporated in the off-diagonal hopping integrals. We have set $Q=(\sqrt{5}+1)/2$ and $\lambda=1$ for both.}
	%\label{trans-fano}
%\end{figure*}
%%%%%%%%%%%%%%%%%%%%%%%%%%%%%%%%%%%%%%%%%%%%%%%%%%%%%

\section{Acknowledgement}
A. M. acknowledges DST for providing her INSPIRE Fellowship $[IF160437]$. A. M. is grateful to Sk Sajid for fruitful discussions.

%\nocite{*}


\begin{thebibliography}{99}% Produces the bibliography via BibTeX.
\bibitem{cayley} A. Cayley, Desiderata and suggestions: No. 2. The Theory of
groups: graphical representation, American Journal of Mathematics \textbf{2}, 1 (1878).
\bibitem{abou} R. Abou-Chacra, D. Thouless, and P. W. anderson, J. Phys. C: Solid State Physics \textbf{6}, 1734 (1973).
\bibitem{efetov} K. Efitov, Zh. Eksp. Teor. Fiz. \textbf{88}, 1032 (1985) [Sov. Phys. JETP \textbf{61}, 606 (1985)].
\bibitem{zirnbauer} M. R. Zirnbauer, Phys. Rev. B \textbf{34}, 6394 (1986).
\bibitem{nakano} M. Nakano, H. Fujita, M. Takahata, and K. Yamaguchi, J. Chem. Phys. \textbf{115}, 1052 (2001).
\bibitem{changlani} H. J. Changlani, S. Ghosh, C. L. Henley, and A. M. L\"{a}uchli, Phys. Rev. B \textbf{87}, 085107 (2013).
\bibitem{lin} C. Lin, Y. Huang, T. Quan, and Y. Zhang, Sc. Rep. \textbf{8}, 15666 (2018).
\bibitem{altshuler} B. L. Altshuler, Y. Gefen, A. Kamenev, and L. S. Levitov, Phys. Rev. Lett. \textbf{78}, 2803 (1997).
\bibitem{mirlin} A. D. Mirlin and Y. V. Fyodorove, Phys. Rev. B \textbf{56}, 13393 (1997).
\bibitem{georgeot} B. Georgeot and D. L. Shepelyanski, Phys. Rev. Lett. \textbf{81}, 5129 (1998).

\bibitem{gornyi} I. V. Gornyi, A. D. Mirlin, and D. G. Polyakov, Phys. Rev. Lett.
\textbf{95}, 206603 (2005).
\bibitem{basko} D. Basko, I. Aleiner, and B. Altshuler, Ann. Phys. 321, 1126
(2006).
\bibitem{li} Z. Jian-Li and J. B. Wang, J. Phys. A: Math. Theor. \textbf{48}, 355301 (2015).
\bibitem{mares} J. Mare\v{s}, J. Novont\'{y}, M. \v{S}tefa\v{n}\'{a}k, and I. Jex, Phys. Rev. B \textbf{101}, 032113 (2020).
\bibitem{sonner} M. Sonner, K. S. Tikhonov, and A. D. Merlin, Phys. Rev. B \textbf{96}, 214204 (2017).
\bibitem{savitz} S. Savitz, C. Peng, and G. Refael, Phys. Rev. B \textbf{100}, 094201 (2019).
\bibitem{eckstein} M. Eckstein, M. Kollar, K. Byczuk, and D. Vollhardt, Phys. Rev. B \textbf{71}, 235119 (2005).
\bibitem{aubry} S. Aubry and G. Andr\'{e}, Ann. Israel Phys. Soc. \textbf{3}, 18 (1980).
\bibitem{harper} P. G. Harper, Proceedings of the Physical Society. Section A \textbf{68}, 874 (1955).
\bibitem{yorikawa} H. Yorikawa, J. Phys. Comm. \textbf{2}, 125009 (2018).

\bibitem{sankar1} S. Das Sarma, S. He, and X. C. Xie, Phys. Rev. Lett. \textbf{61}, 2144 (1988).
\bibitem{sankar2} S. Das Sarma, S. He, and X. C. Xie, Phys. Rev. B \textbf{41}, 5544 (1990).
\bibitem{evers2008} F. Evers and A. D. Mirlin, Rev. Mod. Phys. \textbf{80}, 1355 (2008).
\bibitem{janssen} M. Janssen, Int. J. Mod. Phys. B \textbf{8}, 943 (1994).
\bibitem{aoki} H. Aoki, J. Phys. C: Solid State Phys. \textbf{16} L205 (1983). 
\bibitem{rudo} L. J. Vasquez, A. Rodriguez, and R. A. R\"{o}mer, Phys. Rev. B \textbf{78},
195106 (2008).
\bibitem{luck} C. Godreche and J. M. Luck, J. Phys. A: Gen. Phys. \textbf{23}, 3769
(1990).
\bibitem{hiramoto} H. Hiramoto and M. Kohmoto, Phys. Rev. B \textbf{40}, 8225 (1989).
\bibitem{liu} T. Liu, P. Wang, S. Chen, and G. Xianlong, J. Phys. B: At. Mol.
Opt. Phys. \textbf{51}, 025301 (2018).
\bibitem{pingsheng} Z - Q. Zhang and P. Sheng, Phys. rev. B \textbf{49}, 83 (1994).
\bibitem{alexander} S. Alexander, Phys. Rev. B \textbf{27}, 1591 (1983).
\end{thebibliography}
\end{document}